\documentclass[twocolumn,showpacs,preprintnumbers,amsmath,amssymb]{revtex4}

\usepackage{graphicx}
\usepackage{dcolumn}
\usepackage{bm}

\begin{document}

\title{Dynamic Pairing Effects on Low-Frequency Modes 
of Excitation in Deformed Mg Isotopes close to the Neutron Drip Line}

\author{Kenichi Yoshida$^{1}$}
\author{Masayuki Yamagami$^{2}$}
\author{Kenichi Matsuyanagi$^{1}$}
\affiliation{
$^{1}$Department of Physics, Graduate School of Science, Kyoto University, 
Kyoto 606-8502, Japan
\\
$^{2}$Heavy Ion Nuclear Physics Laboratory, RIKEN, Wako, Saitama 351-0198, Japan}%


\begin{abstract}
Low-frequency quadrupole vibrations in deformed  ${}^{36,38,40}$Mg are 
studied by means of the deformed Quasiparticle-RPA based on the 
coordinate-space Hartree-Fock-Bogoliubov formalism. 
Strongly collective  $K^{\pi}=0^{+}$ and $2^{+}$ excitation 
modes (carrying 10-20 W.u.) are obtained at about 3 MeV. 
It is found that dynamical pairing effects play an essential role
in generating these modes. 
It implies that
the lowest $K^{\pi}=0^{+}$ excitation modes are particularly sensitive 
indicators of dynamical pairing correlations in deformed nuclei near the 
neutron drip line. 

\end{abstract}

\maketitle
\noindent{\bf 1. Introduction}

The physics of drip-line nuclei is one of the current frontiers 
in nuclear structure physics. 
The number of unstable nuclei experimentally accessible will 
remarkably increase when the next generation of radioactive ion beam 
facilities start running, 
and we shall be able to study the properties not only of the ground states
but also of low-lying excited states of drip-line nuclei in the medium-mass 
region. 
In order to quest for the new kinds of excitation mode unique to 
unstable nuclei associated with new features such as neutron skins, 
many attempts have been made using the self-consistent RPA based on the 
Skyrme-Hartree-Fock (SHF) method~\cite{ham96, shl03} 
and the Quasiparticle RPA (QRPA) including the pair correlations
~\cite{mat01,hag01, ben02, yam04, ter05}. 
Most of these calculations, however, are restricted to spherical nuclei, 
and low-frequency RPA modes in deformed unstable nuclei remain largely 
unexplored, except for some recent attempts~\cite{urk01, nak05, hag04a}.

In Ref.~\cite{yos05}, we investigated properties of octupole 
excitations built on superdeformed states in neutron-rich sulfur isotopes 
by means of the RPA based on the deformed Woods-Saxon potential 
in the coordinate-mesh representation. 
We found that low-lying states created by excitation of a single neutron 
from a loosely bound low-$\Omega$ state to a high-$\Omega$ resonance state
acquire extremely strong octupole transition strengths 
due to very extended spatial structure of particle-hole wave functions.
We have extended this work to include pairing correlations
self-consistently, and constructed 
a new computer code that carries out the deformed QRPA calculation  
on the basis of the coordinate-space Hartree-Fock-Bogoliubov (HFB) 
formalism~\cite{dob84}. 
In this paper, we report a new result of the deformed QRPA calculation 
for low-frequency soft quadrupole modes with $K^{\pi}=0^{+}$ and $2^{+}$
in ${}^{36,38,40}$Mg close to the neutron drip line.
According to the Skyrme-HFB calculations~\cite{ter97, sto03}, 
these isotopes are well deformed. 
The shell-model calculation~\cite{cau04} also suggests that the ground state 
of ${}^{40}$Mg is dominated by the neutron $2p$-$2h$ components indicating 
breaking of the $N=28$ shell closure.
We have studied properties of soft modes of excitation in these Mg isotopes
simultaneously taking into account the deformed mean-field effects, 
the pairing correlations, and excitations to the continuum,
and made microscopic analysis of the mechanism of generating these 
collective modes. The result of calculation suggests that
the soft $K^{\pi}=0^{+}$  modes are 
particularly sensitive indicators of dynamical pairing correlations 
in deformed nuclei near the neutron drip line. 

\vspace{0.5cm}
\noindent{\bf 2. Method}

\noindent{\bf 2.1. Mean-field calculation}
\vspace{0.3cm}

Assuming axially symmetric deformation,
we solve the HFB equation~\cite{dob84,obe03},
\begin{equation}
\begin{pmatrix}
h^{q}(\boldsymbol{r}\sigma)-\lambda & \tilde{h}^{q}(\boldsymbol{r}\sigma) \\
\tilde{h}^{q}(\boldsymbol{r}\sigma) & -(h^{q}(\boldsymbol{r}\sigma)-\lambda) \end{pmatrix}\!\!
\begin{pmatrix}
\varphi^{q}_{1,\alpha}(\boldsymbol{r}\sigma) \\ \varphi^{q}_{2,\alpha}(\boldsymbol{r}\sigma)
\end{pmatrix}
\!=\! E_{\alpha}\!\!
\begin{pmatrix}
\varphi^{q}_{1,\alpha}(\boldsymbol{r}\sigma) \\ \varphi^{q}_{2,\alpha}(\boldsymbol{r}\sigma)
\end{pmatrix}, \label{eq:HFB1}
\end{equation}
on a two-dimensional lattice 
discretizing the cylindrical coordinates $(\rho, z)$. 
For the mean-field Hamiltonian $h$, we employ the deformed Woods-Saxon 
potential with the parameters used in~\cite{yos05} 
except for the isovector potential strength 
depending on the neutron excess, for which 
a slightly smaller value, 30~MeV, is adopted in order to 
describe $^{40}$Mg as a drip-line nucleus 
in accordance with the Skyrme-HFB calculations~\cite{ter97, sto03}.
The pairing field $\tilde{h}$ is treated self-consistently using 
the surface-type density-dependent contact interaction, 
\begin{equation}
v_{pp}(\boldsymbol{r},\boldsymbol{r}^{\prime})=
V_{0}\Bigl( 1-\dfrac{\varrho^{\mathrm{IS}}(\boldsymbol{r})}{\varrho_{0}} \Bigr)
\delta(\boldsymbol{r}-\boldsymbol{r}^{\prime}), \label{eq:res_pp}
\end{equation}
with $V_{0}=-450$ MeV$\cdot$fm$^{3}$ and $\varrho_{0}=0.16$ fm$^{-3}$,
where $\varrho^{\mathrm{IS}}(\boldsymbol{r})$ denotes the isoscalar density.
We use the lattice mesh size 
$\Delta\rho=\Delta z=0.8$ fm 
and the box boundary condition at  
$\rho_{\mathrm{max}}=10.0$ fm and $z_{\mathrm{max}}=12.8$ fm. 
The quasiparticle energy is cut off at 50 MeV.

\vspace{0.4cm}
\noindent{\bf 2.2. Quasiparticle-RPA calculation}
\vspace{0.3cm}

Using the quasiparticle basis obtained in the previous subsection,  
we solve the QRPA equation in the standard matrix formulation, 
\begin{equation}
\sum_{\gamma \delta}
\begin{pmatrix}
A_{\alpha \beta \gamma \delta} & B_{\alpha \beta \gamma \delta} \\
B^{*}_{\alpha \beta \gamma \delta} & A^{*}_{\alpha \beta \gamma \delta}
\end{pmatrix}
\begin{pmatrix}
f_{\gamma \delta}^{\lambda} \\ g_{\gamma \delta}^{\lambda}
\end{pmatrix}
=\hbar \omega_{\lambda}
\begin{pmatrix}
1 & 0 \\ 0 & -1
\end{pmatrix}
\begin{pmatrix}
f_{\alpha \beta}^{\lambda} \\ g_{\alpha \beta}^{\lambda}
\end{pmatrix} \label{eq:AB1}.
\end{equation}

For residual interactions, we use the Skyrme-type interaction 
for the particle-hole channel,
\begin{equation}
v_{ph}(\boldsymbol{r},\boldsymbol{r}^{\prime})=
\Bigl[ t_{0}(1+x_{0}P_{\sigma})+\dfrac{t_{3}}{6}(1+x_{3}P_{\sigma})\varrho^{\mathrm{IS}}(\boldsymbol{r}) \Bigr]
\delta(\boldsymbol{r}-\boldsymbol{r}^{\prime}), \label{eq:res_ph}
\end{equation}
with $t_{0}=-1100$ MeV$\cdot$fm$^{3}$, $t_{3}=16000$ MeV$\cdot$fm$^{6}, 
x_{0}=0.5$, $x_{3}=1.0$,
$P_{\sigma}$ being the spin exchange operator.
The density-dependent contact interaction (\ref{eq:res_pp})
is used for the particle-particle channel.
Because the deformed Wood-Saxon potential is used for the mean-field, 
we renormalize the residual interaction in the particle-hole channel 
by multiplying a factor $f_{ph}$
to get the spurious $K^{\pi}=1^{+}$  mode
(associated with the rotation) at zero energy
($v_{ph} \rightarrow f_{ph}\cdot v_{ph}$). 
We cut the model space for the QRPA calculation by
$E_{\alpha}+E_{\beta} \leqq 30$MeV which is 
smaller than that for the HFB calculation.
Accordingly, we need another self-consistency factor $f_{pp}$
for the particle-particle channel. 
We determine this factor such that the spurious $K^{\pi}=0^{+}$ modes 
(associated with the number fluctuation) appear at zero energy
($v_{pp} \rightarrow f_{pp}\cdot v_{pp}$). 

The intrinsic matrix elements $\langle \lambda|Q_{2K}|0 \rangle$ 
of the quadrupole operator $Q_{2K}$
between the excited state $|\lambda \rangle$ 
and the ground state $|0\rangle$ are given by
\begin{equation}
\langle \lambda|Q_{2K}|0 \rangle=\sum_{\alpha \beta}
\Big(Q_{2K}^{\alpha \beta}f_{\alpha \beta}^{\lambda}
+Q_{2K}^{\beta \alpha}g_{\alpha \beta}^{\lambda}\Big)
=\sum_{\alpha \beta}M_{2K}^{\alpha \beta} 
\label{eq:matrix_element},
\end{equation}
where
\begin{equation}
Q_{2K}^{\alpha \beta}
\equiv 2\pi\delta_{K,\Omega_{\alpha}+\Omega_{\beta}}
\int\!\!\mathrm{d}\rho\mathrm{d}z Q_{2K}^{\alpha \beta}(\rho,z). 
\label{eq:integrand} 
\end{equation} 
We calculate the intrinsic strength functions, 
\begin{equation}
S^{\mathrm{IS}}(\omega)
=\sum_{\lambda}|\langle \lambda|Q^{\mathrm{IS}}_{2K}|0 \rangle|^{2} 
\delta(\hbar\omega-\hbar\omega_{\lambda}),
\end{equation}
for the isoscalar quadrupole operators $Q^{\mathrm{IS}}_{2K}$,
and use the notation $B(Q^{\mathrm{IS}}2)=
|\langle \lambda|Q^{\mathrm{IS}}_{2K}|0 \rangle|^{2}$.

\vspace{0.5cm}
\noindent{\bf 3. Results and discussion}
\vspace{0.3cm}

\begin{figure}[tp]
\begin{center}
\includegraphics[scale=0.69]{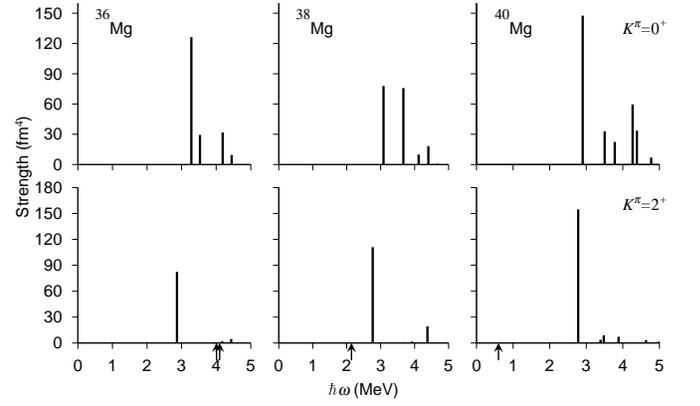}
\caption{Isoscalar quadrupole strengths $B(Q^{\mathrm{IS}}$2) for the 
$K=0^{+}$ (upper panel) and $K=2^{+}$ excitations (lower panel) 
on the prolate ground states of ${}^{36,38,40}$Mg,
obtained by the deformed QRPA calculation with $\beta_{2}=0.28$.
The arrows indicate the neutron threshold energy; 
$E_{\mathrm{th}}=4.01$MeV (1qp continuum), 
4.1 MeV (2qp continuum) for ${}^{36}$Mg, 
2.14 MeV for ${}^{38}$Mg and 0.60 MeV for ${}^{40}$Mg. 
}
\label{Mg_strength}
\end{center}
\end{figure}
\begin{figure}[bp]
\begin{center}
\includegraphics[scale=0.46]{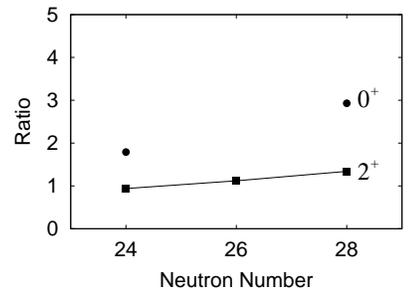}
\caption{Ratios of the neutron and proton matrix elements
$M_{\nu}/M_{\pi}$, divided by $N/Z$, 
are plotted as functions of neutron number. 
Filled circles and squares indicate the ratios for
the lowest $K^{\pi}=0^{+}$ and $2^{+}$ modes, respectively.}
\label{Ratio}
\end{center}
\end{figure}

Calculated quadrupole transition strengths for the 
$K^{\pi}=0^{+}$ and $2^{+}$ excitations in ${}^{36,38,40}$Mg 
are displayed in Fig.~\ref{Mg_strength}. 
We see prominent peaks at 2.9 and 3.3 MeV associated with
the $K^{\pi}=2^{+}$ and $0^{+}$ modes, respectively,
in ${}^{36}$Mg;
at 2.8 MeV for the $K^{\pi}=2^{+}$ mode in ${}^{38}$Mg; 
at 2.8 and 2.9 MeV for the $K^{\pi}=2^{+}$ and $0^{+}$ modes, respectively,
in ${}^{40}$Mg. 
This figure indicates that the transition strengths are
very large (10-20 W.u.) and increase 
as we approach the neutron drip line. 
Figure \ref{Ratio} shows the ratios of the matrix elements for neutrons 
and protons.  
It is seen that the neutron excitation becomes dominant 
as the neutron number increases. 

\begin{figure}[tp]
  \begin{center}
    \begin{tabular}{cc}
	\includegraphics[height=6.1cm]{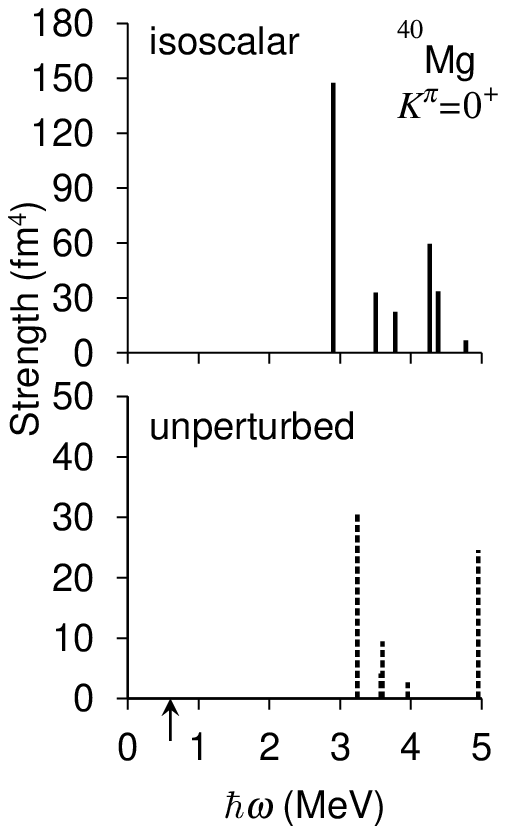}
	\includegraphics[height=6.1cm]{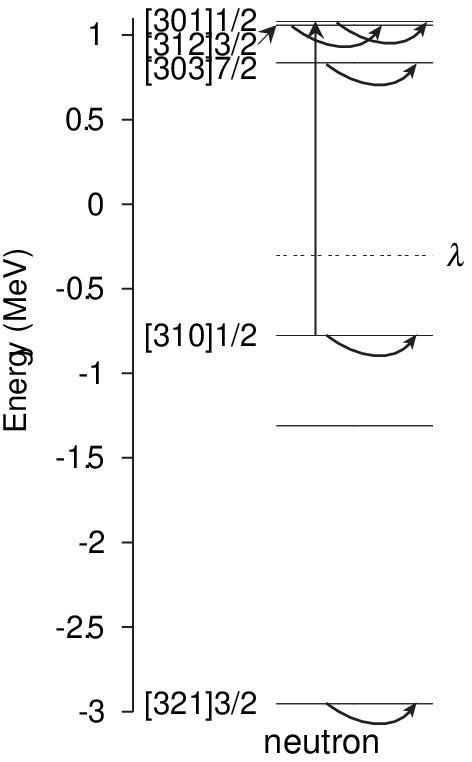}
    \end{tabular}
    \caption{{\it Left}: 
Isoscalar quadrupole strengths $B$(Q$^{\mathrm{IS}}2$) 
for the $K^{\pi}=0^{+}$ excitations in ${}^{40}$Mg are plotted
in the upper panel, while unperturbed two-quasiparticle strengths 
are shown in the lower panel.
The arrow indicates the threshold energy $2|\lambda|=0.60$ MeV. 
{\it Right}: 
Two-quasiparticle excitations generating 
the soft $K^{\pi}=0^{+}$ mode at 2.9 MeV. 
The pertinent levels are labeled 
with the asymptotic quantum numbers $[Nn_{3}\Lambda]\Omega$. 
The Fermi surface for neutrons is indicated by the dashed line.
}
\label{40Mg_0+}
  \end{center}
\end{figure}

\begin{table*}[tp]
\caption{
QRPA amplitudes of the soft  $K^{\pi}=0^{+}$ mode at 2.9 MeV in $^{40}$Mg.
This mode has $B(E2)=2.4 ~e^{2}$fm$^{4}$, 
$B(Q^{\nu}$2)=112 ~fm$^{4}$, and $B(Q^{\mathrm{IS}}2)=148$~fm$^{4}$.
The single-particle levels are labeled with
the asymptotic quantum numbers $[Nn_{3}\Lambda]\Omega$
of the dominant components of the wave functions. 
Only components with $|f_{\alpha \beta}| > 0.1$ are listed. 
}
\label{40Mg_amplitude0+}
\begin{center} 
\begin{tabular}{c|c|c|c|c|c|c}
 & $\alpha$ & $\beta$ & $E_{\alpha}+E_{\beta}$ (MeV) & $f_{\alpha \beta}$ 
 & $Q_{20}^{\alpha\beta}$ (fm$^{2}$)
 & $M_{20}^{\alpha\beta}$ (fm$^{2}$) 
\\ \hline \hline
(a) & $\nu$[310]1/2 & $\nu$[310]1/2 & 3.25 & $-0.633$ & 5.55 & $-3.56$ \\
(b) & $\nu$[312]3/2 & $\nu$[312]3/2 & 3.58 & 0.560 & $-2.04$ & $-1.17$ \\
(c) & $\nu$[301]1/2 & $\nu$[310]1/2 & 3.60 & 0.437 & $-3.08$ & $-1.35$ \\ 
(d) & $\nu$[301]1/2 & $\nu$[301]1/2 & 3.96 & 0.123 & 1.64 & 0.236 \\
(e) & $\nu$[303]7/2 & $\nu$[303]7/2 & 5.01 & 0.277 & $-3.34$ & $-0.987$ \\
(f) & $\nu$[321]3/2 & $\nu$[321]3/2 & 6.97 & $-0.124$ & 3.08 & $-0.423$ 
\end{tabular}
\end{center} 
\end{table*}

\begin{figure}[tp]
\begin{center}
\includegraphics[scale=0.67]{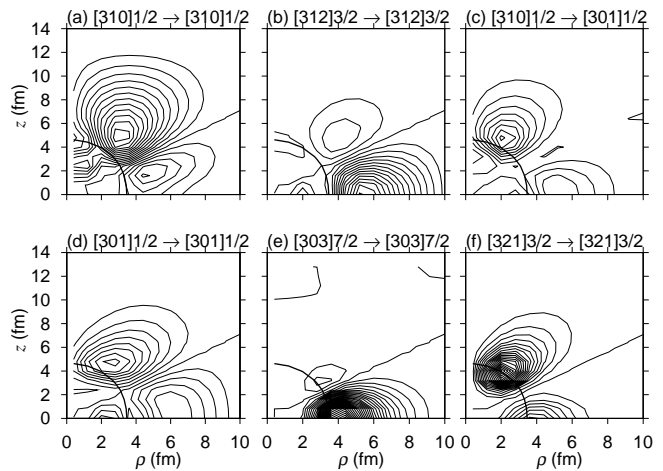}
\caption{Spatial distribution functions $Q_{20}^{\alpha \beta}(\rho,z)$ 
for some two-quasiparticle excitations 
generating the soft $K^{\pi}=0^{+}$ mode in $^{40}$Mg.  
The contour lines are plotted at intervals of 0.002. 
The thick solid line indicates the neutron half density radius.}
\label{integrand}
\end{center}
\end{figure}

\begin{figure}[tp]
\begin{center}
\includegraphics[scale=0.84]{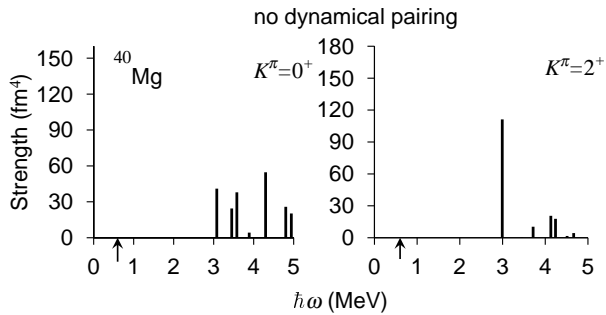}
\caption{Isoscalar quadrupole strengths for the $K^{\pi}=0^{+}$ 
and $2^{+}$ excitations in $^{40}$Mg, obtained 
by switching off the residual particle-particle interactions.}
\label{no_dynamical}
\end{center}
\end{figure}

Let us focus our attention on the properties of excited states 
in ${}^{40}$Mg. 
Details of the transition strength for the $K^{\pi}=0^{+}$ excitations 
are displayed in Fig.~\ref{40Mg_0+}. 
We immediately notice that the transition strength 
for the lowest excited state is significantly enhanced 
from that of the unperturbed two-quasiparticle excitations. 
From the QRPA amplitude listed in Table~\ref{40Mg_amplitude0+}, 
it is clear that this collective mode is generated by coherent superposition 
of neutron excitations of both
particle-hole and particle-particle types.
Let us discuss the reason why this mode acquires eminently large 
transition strength. 
In Ref.~\cite{yos05}, we pointed out several examples where
a neutron excitation from a loosely bound hole state to 
a resonant particle state brings about very large transition strength.
This is a natural consequence of the fact that 
their wave functions are significantly extended from the nuclear surface.
In the present calculation for $^{40}$Mg, this effect is much
enhanced due to the pairing correlations among these loosely bound 
and resonance states.
To show this point, we plot in Fig.~\ref{integrand} 
spatial distributions of two-quasiparticle wave functions 
generating the soft $K^{\pi}=0^{+}$ mode.  
We see that these are significantly extended beyond the half-density radius.
In addition, their spatial structures are rather similar; 
this is a favorable situation to generates coherence among them. 
In Fig.~\ref{no_dynamical} we show how the transition strengths 
for the $K^{\pi}=0^{+}$ and $2^{+}$ modes change 
if the residual particle-particle interactions are ignored. 
We see that, 
although the prominent peak for the $K^{\pi}=2^{+}$ mode still remains, 
the transition strength decreases.
For the $K^{\pi}=0^{+}$ mode, we notice a more striking effect;
the prominent peak seen in Fig.~\ref{Mg_strength} 
disappears when the dynamical pairing correlation is absent. 
Therefore we can say that the coherent superposition of particle-hole, 
particle-particle and hole-hole excitations
is indispensable for the emergence of this mode.
The importance of the coupling between 
the (particle-hole type) $\beta$ vibration and 
the (particle-particle and hole-hole type) pairing vibration 
has been well known in stable deformed nuclei~\cite{BM2}.
A new feature of the soft $K^{\pi}=0^{+}$ mode in neutron drip-line
nuclei is that this coupling takes place among two-quasiparticle excitations
that are loosely bound or resonances, so that the transition strengths are
extremely enhanced. 

\begin{figure}[tp]
\begin{center}
\includegraphics[scale=0.84]{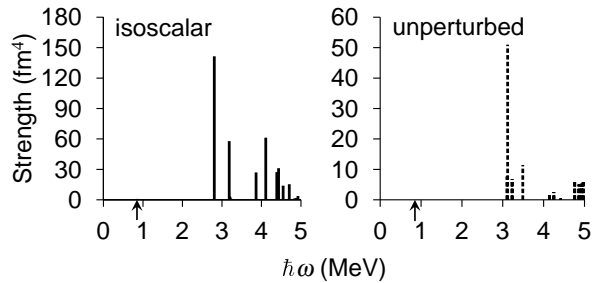}
\caption{{\it Left}: The isoscalar transition strengths 
for the $K^{\pi}=0^{+}$ excitations in $^{40}$Mg 
calculated using the box with size 
$\rho_{\mathrm{max}} \times z_{\mathrm{max}} = 13.2$ fm $\times 16.0$ fm. 
{\it Right}: Unperturbed two-quasiparticle strengths.}
\label{strength_large}
\end{center}
\end{figure}
We examined the numerical stability of this soft $K^{\pi}=0^{+}$ mode 
against variation of the box size. 
In Fig.~\ref{strength_large} 
we show the isoscalar and unperturbed two-quasiparticle transition strengths 
for the $K^{\pi}=0^{+}$ excitations in $^{40}$Mg,
obtained by using a larger box; 
$\rho_{\mathrm{max}} \times z_{\mathrm{max}} = 
13.2$ fm $\times 16.0$ fm. 
We see that the position and the strength of the prominent peak 
are almost unchanged. 
This result confirms that the main two-quasiparticle components 
generating this mode are weakly bound and resonant states, and that
the numerical error associated with the continuum discretization
is not very significant.

\newpage
\vspace{0.5cm}
\noindent{\bf 4. Conclusion}
\vspace{0.3cm}

We have investigated properties of excitation modes 
in deformed Mg isotopes close to the neutron drip line 
by means of the deformed QRPA  based on the coordinate-space HFB.
For ${}^{36,38,40}$Mg,
we have obtained the soft $K^{\pi}=0^{+}$ and $2^{+}$ modes possessing
large transition strengths.
The microscopic mechanism of enhancement of 
the $K^{\pi}=0^{+}$ strength is especially interesting. 
There are two causes for this enhancement;
1) the strong coupling occurs between the quadrupole shape vibration and 
the neutron pairing vibration, and
2) the two quasiparticle configurations coherently generating the
$K^{\pi}=0^{+}$ modes possess significantly extended spatial structures.
The first cause has been well known 
in studies of stable deformed nuclei~\cite{BM2}, 
but the second cause is unique to unstable deformed nuclei 
near the neutron drip line.
The nature and roles of pair correlations for collective excitations 
in drip-line nuclei is currently under lively discussions~\cite{mat05}.
On the basis of the deformed QRPA calculation demonstrating that 
the soft $K^{\pi}=0^{+}$ mode is quite sensitive to the dynamical pairing,
we suggest that an appearance of this soft mode is a very good indicator 
of pairing correlations in deformed drip-line nuclei. 
\vspace{0.5cm}

\noindent{\bf Acknowledgments}
\vspace{0.5cm}

This work was done as a part of the Japan-U.S. Cooperative Science Program
``Mean-Field Approach to Collective Excitations in Unstable
Medium-Mass and Heavy Nuclei", 
and is supported by a
Grant-in-Aid for the 21st Century COE ``Center for Diversity and
Universality in Physics" from the Ministry of Education, Culture, Sports,
Science and Technology (MEXT) of Japan, 
and also 
by a Grant-in-Aid  for Scientific Research No.~16540249  
from the Japan Society for the Promotion of Science.
The numerical calculations were performed on the NEC SX-5 supercomputer
at Yukawa Institute for Theoretical Physics, Kyoto University.


\end{document}